\documentclass[aps,reprint,prd,showpacs,amsmath,amssymb,nofootinbib]{revtex4-1}
\usepackage{graphicx}
\usepackage{color}
\usepackage{times}
\usepackage{inputenc}
\usepackage{bm}
\usepackage{multirow}
\usepackage{float}
\usepackage{url}
\usepackage{natbib}
\usepackage{tensor}
\usepackage[colorlinks=true,citecolor=blue,urlcolor=blue,linkcolor=red]{hyperref}
\usepackage[normalem]{ulem}

\def\msun{\,M_{\odot}}
\def\fm3{\;\text{fm}^{-3}}
\def\mev{\;\text{MeV}}
\def\gev{\;\text{GeV}}

\newcommand{\black}[1]{{\color[rgb]{0.0,0.0,0.0}#1}}
\newcommand{\green}[1]{{\color[rgb]{0.0,0.7,0.0}#1}}

\begin{document}

\title{
Constraining the competition between the deconfinement and chiral phase transitions in light of the multimessenger era}
\author{Wen-Li Yuan$^{1}$}
\email{wlyuan@pku.edu.cn}
\author{Bikai Gao$^{2,3}$}
\email{gaobikai@hken.phys.nagoya-u.ac.jp}
\author{Yan Yan$^{3}$} 
\email{2919ywhhxh@163.com}
\author{Bolin Li$^{4}$}
\email{blli@usst.edu.cn}
\author{Renxin Xu$^{1}$}
\email{r.x.xu@pku.edu.cn} 

\affiliation{$^1$School of Physics and State Key Laboratory of Nuclear Physics and Technology, Peking University, Beijing 100871, China;\\ 
$^2$Department of Physics, Nagoya University, Nagoya 464-8602, Japan;\\
$^3$Research Center for Nuclear Physics (RCNP), Osaka University, Osaka 567-0047, Japan;\\
$^4$School of Microelectronics and Control Engineering, Changzhou University, Changzhou 213164, China;\\
$^5$ Department of Physics, University of Shanghai for Science and Technology, Shanghai 200093, China
}

\date{\today}

\begin{abstract}

We extend the parity doublet model for hadronic matter and study the possible presence of quark matter inside the cores of neutron stars with the modified Nambu-Jona-Lasinio (NJL) model.  Considering the uncertainties of the QCD phase diagram and the location of the critical endpoint, we aim to explore the competition between the chiral phase transition and the deconfinement phase transition systematically, regulated by the vacuum pressure $-B$ in the NJL model. Employing a Maxwell construction, a sharp first-order deconfinement phase transition is implemented, combining the parity doublet model (PDM) for the hadronic phase and the NJL model for the \black{high density} quark phase. The position of chiral phase transition is obtained from the NJL model self-consistently. We find stable neutron stars with a quark core within a specific parameter space that satisfies current astronomical observations. The observations suggest a relatively large chiral invariant mass $m_0=600\mev$ in the PDM and a larger split between the chiral and deconfinement phase transitions while assuming the first-order deconfinement phase transition. The maximum mass of the hybrid star that we obtain is $\sim 2.2\msun$. 

\end{abstract}

\maketitle 
\section{Introduction}
The study of quantum chromodynamics (QCD) phase transitions is a crucial topic in hadron physics, with numerous experimental and theoretical programs dedicated to this field. A central goal of ultrarelativistic heavy-ion collision experiments is to understand the quark-hadron phase transition. As temperature and density increase, strongly interacting matter is expected to transition from hadronic phase to quark-gluon plasma~\cite{2005PhR...407..205B,2008PhRvD..77k4028F,2011RPPh...74a4001F,2017EPJWC.14104001L}. However, the existence and location of the critical endpoint in the QCD phase diagram remain open questions. In addition, the relationship between the deconfinement phase transition and the chiral phase transition is not yet fully understood.  Some studies have considered these two types of transitions to occur simultaneously in the framework of QCD phenomenological models, such as Refs.~\cite{2007PhRvD..76g4023S,2015PhRvC..91c5803L,2017ApJ...836...89K,2018PhRvD..98h3013L,2020ApJ...904..103M,2022JPhG...49d5201L,2022ApJ...935...88H,2023ApJ...944..206L}. Despite lattice results showing that both transitions occur at the same temperature and zero baryon density~\cite{2002LNP...583..209K,2003ARNPS..53..163L}, at finite baryon chemical potential, this coincidence is an open question~\cite{2008PhRvD..77k4028F,2011RPPh...74a4001F}. Recent studies suggest the possible existence of a quarkyonic phase, characterized by 
chiral symmetric while quarks remain confined~\cite{2007NuPhA.796...83M,2011PhLB..696...58H,2011PhRvD..84c4028S,2012NuPhA.877...70K}. The presence of such a phase in the QCD phase diagram could have significant implications for the internal structure of neutron stars, particularly regarding the possibility of quark matter in their cores~\cite{2016ApJ...817..180F,2019PhRvL.122l2701M}.

In recent years, the advancement of the multimessenger astronomy era has provided tight constraints on the equation of state (EOS) of dense matter, which are useful tools for accessing the properties of QCD matter. For a comprehensive review of this topic, see, e.g., Refs~\cite{2008RvMP...80.1455A,2018RPPh...81e6902B,2024LRR....27....3K}. In particular, the mass and radius of neutron stars encode unique information on the EOS at supranuclear densities. Several massive pulsars with a mass of about two-solar mass with $M=1.908 \pm 0.016\msun$ (PSR J1614-2230)~\cite{2010Natur.467.1081D}, $M=2.01 \pm 0.04\msun$ (PSR J0348+0432)~\cite{2013Sci...340..448A}, $M=2.14^{+0.10}_{-0.09} \msun$ (PSR J0740+6620)~\cite{2020NatAs...4...72C} detected during the last decade spark discussions about the possibility of quark matter being present at the core of neutron stars, giving us opportunities to study the competition between the deconfinement and chiral phase transitions in natural laboratory. 

In this study, we perform an extensive study of hybrid stars using several parameterizations of a parity doublet model (PDM)~\cite{1989PhRvD..39.2805D,2001PThPh.106..873J,2008PhRvC..77b5803D,2013PhRvC..87a5804D,2015PhRvC..92b5201M,2017PhRvC..95e9903M,2021PhRvC.103d5205M,2021PhRvC.104f5201M,2022PhRvC.106f5205G,2023Symm...15..745M,2024PhRvC.109f5807G,2024arXiv240318214G,Marczenko:2023txe,Marczenko:2024jzn} together with a typical three-flavor NJL model with scalar four-fermion interactions, ’t Hooft six-fermion interactions and the Fierz transformed interactions~\cite{1992RvMP...64..649K,1994PhR...247..221H,2019ChPhC..43h4102W,2020PhRvD.102e4028S,2020PhRvD.101f3023L,2022PhRvD.105l3004Y,2023PhRvD.108d3008Y,2024ApJ...966....3Y}. Within this approach, the hadronic and quark degrees of freedom are derived from different Lagrangians. We employ the PDM for the hadronic phase, in which the nucleon masses not only have a mass associated with the chiral symmetry breaking, but also a chiral invariant mass $m_0$, which is insensitive to the chiral condensate and the presence of which is manifested by recent lattice QCD simulations~\cite{2015PhRvD..92a4503A,2017JHEP...06..034A,2019PhRvD..99g4503A}. For the quark phase, we utilize a modified NJL model that captures the dynamics of spontaneous chiral symmetry breaking. Different from the standard NJL model, which includes the vector repulsion by hand, we use a universal self-consistent mean-field approximation method to introduce the vector interaction through Fierz transformation, then the exchange interactions are further considered weighted by $\alpha$. It is a more general version of the NJL-like model, compared to the specific choice of $\alpha=0.5$ introduced in Ref.~\cite{1992RvMP...64..649K}. As mentioned in Wang et al.~\cite{2019ChPhC..43h4102W}, $\alpha$ should be constrained by finite-density experimental data for the application to strong interaction matter in neutron stars. For the present study of hybrid stars, the $\alpha$ values are constrained by the stability argument of the presence of quark matter and maximum mass observations, and the results of the star properties are discussed in detail under different choices of $\alpha$.

By Maxwell construction, the deconfinement transition is associated with the point where both models have the same free energy, and chiral symmetry restoration in the quark model occurs at a chemical potential $\mu_{\chi}$ where the chiral condensate exhibits the maximum change within the NJL model. This is generally distinct from the chemical potential of deconfinement, $\mu_{\mathrm{de}}$. Although when the current quark mass goes to infinity, QCD becomes pure gauge $\rm SU(3)$ theory, which is center symmetric in the vacuum, and then the Polyakov-loop is usually used to examine the deconfinement phase transition~\cite{2007PhRvD..76g4023S}, it is not satisfying at finite chemical potential~\cite{2001PhRvD..63l1702F,2007NuPhA.796...83M}, and our approach permits a flexible and systematic exploration of the competition between the deconfinement and chiral phase transitions. This methodology may provide deeper insights into the true properties of high-density QCD matter, potentially unveiling new aspects of the QCD phase diagram. In this study, the unknown magnitude of this hypothetical split between them will be studied parametrically by changing the value of the vacuum pressure $P(\mu=0; M)$, which is usually introduced in the grand thermodynamic potential of the NJL model in order to force a vanishing pressure at zero temperature and chemical potential. Instead of calculating this vacuum pressure within the NJL model, we treat $P(\mu=0; M)$ as a phenomenological free parameter corresponding to $-B$ (with $B$ being the vacuum bag constant), which preserves quark confinement.

This paper is organized as follows. In Sec.~\ref{Sec: formulalism}, we will briefly present the hadronic EOS employed in this work, and will introduce the (2+1)-flavor modified NJL models for describing the quark matter, including the Fierz transformed interactions. Section~\ref{Sec: Results} discusses the interplay between the deconfinement and chiral phase transitions, as well as the results on hybrid matter EOS and hybrid stars, along with the observational constraints. Our results are summarized in Sec.~\ref{sec:summary}.

\maketitle 
\section{\green{Formulism} }\label{Sec: formulalism}
\subsection{Hadronic matter within PDM model} \label{Sec:hadronic matter}
The EOS of nuclear matter obtained within the PDM has been amply discussed in previous works~\cite{2015PhRvC..92b5201M,2017PhRvC..95e9903M,2021PhRvC.103d5205M,2021PhRvC.104f5201M,2022PhRvC.106f5205G,2023Symm...15..745M,2024PhRvC.109f5807G}. In this approach, the positive and negative parity states of the nucleons are treated as chiral partners, and their masses become degenerate when chiral symmetry is restored at high densities. This mass, known as the chiral invariant mass, is a key parameter in these models, significantly influencing the stiffness of the EOS. In particular, a larger $m_0$ results in weaker $\sigma$ couplings to nucleons since a nucleon's mass is not completely derived from the $\sigma$ fields. Correspondingly, the couplings to $\omega$ fields are reduced since at the nuclear saturation density $n_0$, the repulsive contributions of the $\omega$ fields must be counterbalanced by the attractive $\sigma$ contributions. Beyond densities greater than $n_0$, the $\sigma$ fields decrease while the $\omega$ fields increase, leading to an imbalance. As a consequence, a larger $m_0$ weakens the $\omega$ fields and softens EOS at supranuclear densities. Typical PDMs are $\sigma$-$\omega$ type mean field models, with some works also incorporating the isovector scalar meson $a_0(980)$, which is believed to appear in asymmetric matter, such as in neutron stars, and affect the nuclear symmetry energy coefficients~\cite{2025arXiv250616684K}. However, as investigated in Ref.~\cite{2023PhRvC.108e5206K}, the inclusion of the $a_0(980)$ has a negligible impact on the macroscopical mass-radius properties of neutron stars, resulting in only a slight increase in the radius of less than $1 \rm km$. In this study, we then consider the PDM model with $N_f=2$ and include the vector meson mixing, such as the $\omega^2\rho^2$ interaction, as described in Ref.~\cite{2024PhRvC.109f5807G}. 

The thermodynamic potential of the model in the mean-field approximation is calculated as
\begin{equation}
\begin{aligned}
\Omega_{\mathrm{PDM}}&=V_\sigma-V\left(f_{\pi}\right)\\
&+V_\omega+V_\rho + V_{\omega\rho} +\sum_{i=+,-} \sum_{x=p, n} \Omega_{x} \ ,\label{Eq:Omega PDM}
\end{aligned}
\end{equation}
where $i = +, -$ denote the positive-parity ordinary nucleon $N(939)$ and negative-parity excited nucleon $N^{*}(1535)$.
The mean-field potential $V(\sigma)$, $V_\omega$, $V_\rho$ and $V_{\omega \rho}$ are given by
\begin{equation}
\begin{aligned}
V(\sigma) = -\frac{1}{2}\bar{\mu}^{2}\sigma^{2} &+ \frac{1}{4}\lambda_4 \sigma^4 -\frac{1}{6}\lambda_6\sigma^6 - m_{\pi}^{2} f_{\pi}\sigma\ \ , \\
V_\omega&=-\frac{m_\omega^2}{2} \omega^2 \ ,\\
V_\rho&=-\frac{m_\rho^2}{2} \rho^2 \\
V_{\omega \rho}&=-\lambda_{\omega\rho}(g_{\omega NN}\omega)^2(g_{\rho NN}\rho)^2\ ,\label{Eq:PDM potential}
\end{aligned}
\end{equation}
with $f_{\pi}$ the pion decay constant. Here, $\bar{\mu}, \lambda_{4}, \lambda_{6}$ and $\lambda_{\omega\rho}$ are parameters to be determined and the kinetic part of the thermodynamic potential $\Omega_x$ reads
\begin{equation}
\begin{aligned}
\Omega_x= -2 \int^{k_x^{\pm}} \frac{\mathrm{d}^3 \mathbf{p}}{(2 \pi)^3}\left(\mu_x^*-E_{\mathbf{p}}^i\right),\ \label{Eq: PDM kinetic part} 
\end{aligned}
\end{equation}
with $E_{{\bf p}}^{i} = \sqrt{{\bf p}^{2} + m_{\pm}^{2}}$ is the energy of relevant nucleon with mass $m_{\pm}$ and momentum ${\bf p}$, and $k_{x}^{\pm}=\sqrt{(\mu^{*}_{x})^2-m_{\pm}^2}$ is the fermi momentum for the relevant particle, in which $\mu_x^{*}$ is the effective chemical potential. We notice that we use the no-sea approximation, assuming that the structure of the Dirac sea remains the same for the vacuum and medium. 

With the mirror assignment of chirality, one can construct a chiral-invariant mass term $m_0$. The masses of the positive- and negative-parity chiral partners then arise from two different mechanisms: the mass generated by the chiral condensate and the chiral invariant mass $m_0$, which are expressed as follows:
\begin{equation}
\begin{aligned}
m_{ \pm}=\frac{1}{2}\left[\sqrt{\left(g_1+g_2\right)^2 \sigma^2+4\left(m_0\right)^2} \mp\left(g_1 - g_2\right) \sigma\right]\ ,\label{Eq: PDM mass}
\end{aligned}
\end{equation}
where $\pm$ sign denotes parity. The spontaneous chiral symmetry breaking yields the mass splitting between the two baryonic parity partners in each parity doublet. When the symmetry is restored, the masses in each parity doublet become degenerate: $m_{\pm}(\sigma=0)=m_0$. The positive-parity nucleons are identified as the positively charged and neutral $N(939)$ states: proton $(p)$ and neutron $(n)$. Their negative-parity counterparts, denoted as $p^{*}$ and $n^{*}$, are identified as $N(1535)$ resonance. For a given chirally invariant mass, $\mathrm{m}_0$, the parameters $\mathrm{g}_1$ and $\mathrm{g}_2$ are determined by the corresponding vacuum masses, $m_N=939\mev,  m_{N^*}=1500\mev$. The effective chemical potentials for nucleons and their chiral partners are given by
\begin{equation}
\begin{aligned}
\mu_p=\mu_p^{*}&=\mu_Q+\mu_B-g_{\omega NN} \omega- \frac{1}{2}g_{\rho NN} \rho \ , \\
\mu_n=\mu_n^{*}&=\mu_B - g_{\omega NN} \omega+ \frac{1}{2}g_{\rho NN} \rho \ .\label{Eq: PDM chemicalPoten}
\end{aligned}
\end{equation}
 
The total thermodynamic potential of the hadronic matter in neutron stars is obtained by including the effects of leptons as
\begin{equation}
\begin{aligned}
\Omega_{\mathrm{H}}=\Omega_{\mathrm{PDM}} + \Omega_{e}\ ,
\end{aligned}
\end{equation}
where $\Omega_{e}$ is the thermodynamic potentials for electrons given by 
\begin{equation}
\begin{aligned}
\Omega_{e}=-2 \int^{k_F} \frac{\mathrm{d}^3 \mathbf{p}}{(2 \pi)^3}\left(\mu_e-E_{\mathbf{p}}^e\right)\ ,
\end{aligned}
\end{equation}
Finally, we have the pressure in hadronic matter as
\begin{equation}
P_{\mathrm{H}}=-\Omega_{\mathrm{H}}\ .
\end{equation}
Using the explicit parameter sets determined in Ref.~\cite{2024PhRvC.109f5807G}, with fitting to the pion decay constant and hadron masses, as well as to the normal nuclear matter properties, we can calculate the corresponding EOS in the hadronic phase for different choices of the chiral invariant mass $m_0$.

We note that, throughout this work, we assume that the hyperons do not enter into the system in hadronic phase and consider the case $N_f=2$. For three-flavor PDM, one would need to include the parity doublets of hyperons ($\Lambda$, $\Sigma$, $\Xi$) with their negative parity partners, significantly expanding the parameter space and computational complexity.

\subsection{Quark matter within (2+1)-flavor model}\label{Sec: NJL model}
In this section, we introduce the (2+1)-flavor NJL model to describe the effective interactions between quarks. The Lagrangian density of (2+1)-flavor NJL model is
\begin{equation}
\begin{aligned}
\mathcal{L}_{\mathrm{NJL}}^{~3f} &=\mathcal{L}_{0}+\mathcal{L}_{\mathrm{int}}^{~3f} \ , \\
\mathcal{L}_{\mathrm{int}}^{~3f} &=  \mathcal{L}_{\sigma}^{4}+ \mathcal{L}_{\sigma}^{6}  \ ,
\label{eqNJL}
\end{aligned}
\end{equation}
in which the phenomenological scalar interaction term $\mathcal{L}_{\sigma}^{4}$ is 
\begin{equation}
\mathcal{L}_{\sigma}^{4} =\sum_{i=0}^{8} G\left[\left(\bar{\psi} \lambda_{i} \psi\right)^{2}+\left(\bar{\psi} i\gamma^{5}\lambda_{i} \psi\right)^{2}\right] \ . \label{eqNJLscalar}
\end{equation}
The six-fermion interaction term $\mathcal{L}_{\sigma}^{6} $ is written as: 
\begin{equation}
\mathcal{L}_{\sigma}^{6} = -K\left(\operatorname{det}\left[\bar{\psi}\left(1+\gamma^{5}\right) \psi\right]+\operatorname{det}\left[\bar{\psi}\left(1-\gamma^{5}\right) \psi\right]\right) \ . \label{eqNJLsixfermion}
\end{equation}
It represents the effects of the instanton-induced QCD axial anomaly, which is a determinant in flavor space and breaks the $U(1)_{A}$ axial symmetry of the QCD Lagrangian.
In NJL model, $G$ and $K$ are the four-fermion and six-fermion interaction coupling constants, respectively. $\lambda_{i}\; (i=1 \rightarrow 8)$ is the Gell-Mann matrix in flavor space. $\lambda_{0}= \sqrt{2/3}\; I_{0}$ ($I_{0}$ is the identity matrix).

In the following, we further consider the effect of a rearrangement of fermion field operators by applying the Fierz transformation to the interaction terms in the (2+1)-flavor NJL model~\cite{2019ChPhC..43h4102W,2020PhRvD.102e4028S,2020PhRvD.101f3023L,2022PhRvD.105l3004Y,2024ApJ...966....3Y,2023PhRvD.108d3008Y} as follows:
\begin{equation}
\mathcal{F}(\mathcal{L}_{\mathrm{int}}^{3f} )=\mathcal{F}(\mathcal{L}_{\sigma}^{4})+ \mathcal{F}(\mathcal{L}_{\sigma}^{6}) \ .\label{eqFierz}
\end{equation}
The Fierz identity of the four-fermion scalar and pseudoscalar interaction term $\mathcal{F}(\mathcal{L}_{\sigma}^{4})$, considering only the contributions of color-singlet terms, is
\begin{equation}
\mathcal{F}(\mathcal{L}_{\sigma}^{4})=-\frac{ G}{2}\left[\left(\bar{\psi} \gamma_{\mu} \lambda_{i}^{0} \psi\right)^{2}-\left(\bar{\psi} \gamma_{\mu} \gamma_{5} \lambda_{i}^{0} \psi\right)^{2}\right] \ .\label{eqFierzscalar}
\end{equation}
Since this Fierz transformation of six-fermion interaction can be defined as a transformation which leaves the interaction invariant under all possible permutations of the quark spinors $\psi$ occurring in it~\cite{1992RvMP...64..649K}, the six-fermion interaction term does not change after the Fierz transformation: 
\begin{equation}
\mathcal{F}(\mathcal{L}_{\sigma}^{6})=\mathcal{L}_{\sigma}^{6} \ .\label{eqFierzSix}
\end{equation}
Due to the mathematical equality between the original interactions and Fierz transformed interactions, we can combine them using a weighting factor $\alpha$. Then the effective Lagrangian becomes:
\begin{equation}
\mathcal{L}_{\rm eff}^{~3f}=\bar{\psi}(i \gamma^{\mu}\partial_{\mu} -m+\mu\gamma^{0}) \psi +(1-\alpha)\mathcal{L}_{\mathrm{int}}^{~3f}+\alpha \mathcal{F}(\mathcal{L}_{\mathrm{int}}^{~3f}) \ .\label{eqNJLeff}
\end{equation}

Under the mean-field approximation, the mass gap equations and the effective chemical potential $\mu_{f}^{*}$ of flavor $f$ can be obtained as follows: 
\begin{equation}
\begin{aligned}
M_{f}&= m_{f}-4 (1-\alpha) G \sigma_{f}+2 K \sigma_{j} \sigma_{k}\\
&= m_{f}-4G'\sigma_{f}+2 K \sigma_{j} \sigma_{k} , \label{eq:Effmass}
\end{aligned}
\end{equation}
\begin{equation}
\begin{aligned}
\mu_{f}^{*}&=\mu_{f} -\frac{2}{3}\alpha G\sum_{f^{\prime}=u, d, s}\rho_{f^{\prime}}\ .\label{eqGapEq}
\end{aligned}
\end{equation}
where we define $G'=(1-\alpha)G$ and $f, j, k $ are the even permutations of $u, d, s$. From Eqs.~(\ref{eq:Effmass}-\ref{eqGapEq}), it is clear that the introduction of Fierz transformed identity contributes to the renormalized chemical potential and the gap equation. 

\begin{figure*}
\centering
{\includegraphics[width=0.49\textwidth]{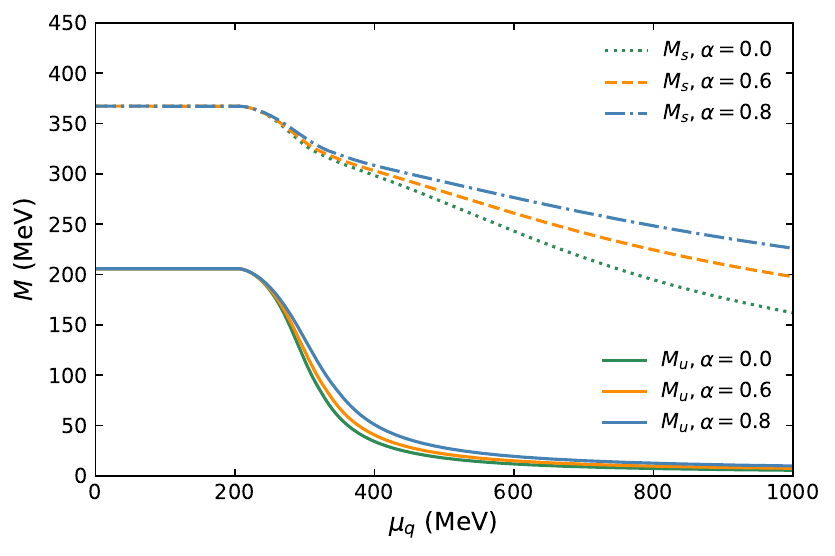}}
{\includegraphics[width=0.49\textwidth]{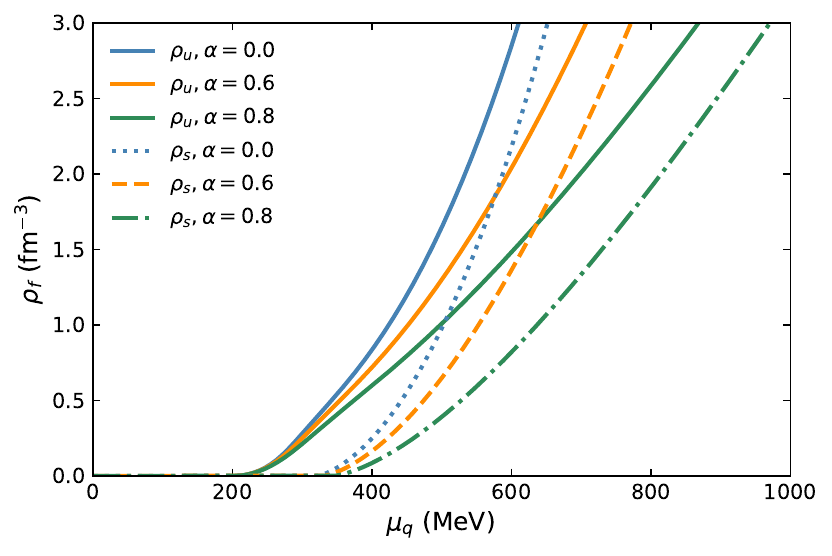}}
\caption{Upper penal: Dynamical quark mass $M$ of $u$, $d$ and $s$ quarks versus quark chemical potential $\mu_q$ for modified (2+1)-flavor NJL models with $\alpha=0.0$, $\alpha=0.6$ and $\alpha=0.8$. Lower penal: The corresponding quark number densities $\rho_f$ as functions of $\mu_q$ with the same parameter sets.}
\label{fig:mass_rho}
\end{figure*}

In the following, we focus on how to obtain the crucial quantities of quark condensate and the quark number density, as well as present the regularization procedure we used. In Euclidean space, introducing the chemical potential at zero temperature is equivalent to performing a transformation~\cite{1994PhR...247..221H,2005PhR...407..205B,2005PhRvC..71a5205Z}:
$p_{4} \rightarrow p_{4}+i \mu_{f}^{*}$.
Thus integrate over $p_{4}$ first and apply the following proper-time regularization,
\begin{equation}
\begin{aligned}
&\frac{1}{A^{n}}=\frac{1}{(n-1) !} \int_{0}^{\infty} \mathrm{d} \tau \tau^{n-1} e^{-\tau A} \\
&\stackrel{\text { UV cutoff }}{\longrightarrow} \frac{1}{(n-1) !} \int_{\tau_{\mathrm{UV}}}^{\infty} \mathrm{d} \tau \tau^{n-1} e^{-\tau A} \ ,  \label{eq:NJL PRT}
\end{aligned}
\end{equation}
where the lower cutoff $\tau_{\rm UV}=1/\Lambda_{\rm UV}^2$ induces the dumping factor into the original propagator, dumping high frequency contribution~\cite{2005PhR...407..205B,2020PhRvD.102e4028S,2020PhRvD.101f3023L,2022PhRvD.105l3004Y}, and $\Lambda_{\rm UV}$ is the parameter related to ultraviolet cutoff. Then, one can obtain the analytical results of quark condensate as follows:
\begin{equation}
\begin{aligned}
&\sigma_{f}=
-N_{\mathrm{c}} \int_{-\infty}^{+\infty} \frac{\mathrm{d}^{4} p^{\mathrm{E}}}{(2 \pi)^{4}} \frac{4  M_{f}}{\left(p^{\mathrm{E}}\right)^{2}+M_{f}^{2}}\\
&=-N_{\mathrm{c}} \int_{-\infty}^{+\infty} \frac{\mathrm{d}^{4} p^{\mathrm{E}}}{(2 \pi)^{4}} \frac{4  M_{f}}{\left(  p_{4}+ i\mu_{f}^{*} \right)^{2}+ p^{2}+M_{f}^{2}}
\\
&= \begin{cases}-\frac{3 M_{f}}{\pi^{2}} \int_{\sqrt{\mu_{f}^{* 2}-M_{f}^{2}}}^{+\infty} \mathrm{d} p \frac{\left[1-\operatorname{Erf}\left(\sqrt{M_{f}^{2}+p^{2}} \sqrt{\tau_{\mathrm{UV}}}\right)\right] p^{2}}{\sqrt{M_{f}^{2}+p^{2}}},  M_{f}<\mu_{f}^{*} \\ \frac{3 M_{f}}{4 \pi^{2}}\left[M_{f}^{2} \Gamma\left(0, M_{f}^{2}\tau_{\mathrm{UV}} \right) -\frac{e^{-M_{f}^{2} \tau_{\mathrm{Uv}}}}{\tau_{\mathrm{UV}}}\right], M_{f}>\mu_{f}^{*}\end{cases} \label{eqNJLsigma}
\end{aligned}
\end{equation}
where $\Gamma\left(a, z \right)= \int_{z}^{+\infty} \mathrm{d} t \; t^{a-1} e^{-t}$ and $\operatorname{Erf}(x)=\frac{2}{\sqrt{\pi}} \int_{0}^{x} \mathrm{d} t \; e^{-t^{2}}$.

At zero temperature and finite chemical potential, the quark number density is
\begin{equation}
\begin{aligned}
\rho_{f} &=\left\langle\psi^{+} \psi\right\rangle_{f} =-\int \frac{\mathrm{d}^{4} p}{(2 \pi)^{4}} \operatorname{Tr}\left[i S_{f}\left(p^{2}\right) \gamma_{0}\right] \\
&=2 N_{\mathrm{c}} \int \frac{\mathrm{d}^{3} p}{(2 \pi)^{3}} \theta\left(\mu_{f}^{*}-\sqrt{p^{2}+M_{f}^{2}}\right) \\
&= \begin{cases}\frac{1}{\pi^{2}}\left(\sqrt{\mu_{f}^{* 2}-M_{f}^{2}}\right)^{3} \ , & \mu_{f}^{*}>M_{f} \\
0\ . & \mu_{f}^{*}<M_{f}\end{cases}\ . \label{eqNJLQN}
\end{aligned}
\end{equation}
Eq.~(\ref{eqNJLQN}) obviously shows that the quark number density of flavor $f$ will be nonzero when the effective quark chemical potential $\mu^{*}_f$ exceeds a threshold value.

Before doing calculations, we should fix the parameter sets first. At zero temperature and quark chemical potential, apart from $\alpha$, the fixing of the model parameters is the same with the original version of the NJL model~\cite{1994PhR...247..221H}.
According to the latest edition of the Review of Particle Physics Ref.~\cite{2020PTEP.2020h3C01P}, the current quark mass $m_u$ and $m_s$ are predicted to be $\bar{m}=\left(m_u+m_d\right) / 2=3.5_{-0.2}^{+0.5} \mev$ and $m_s=95_{-3}^{+9} \mev$ respectively. Similar to the procedure in Ref.~\cite{1994PhR...247..221H}, after fixing the masses of the up and down quarks by equal values, the other parameters $m_s, \Lambda_{\mathrm{UV}}, G', K$ are chosen to reproduce the experimental data of the pion decay constant and pion mass for $f_{\pi}=92\mev,\,M_{\pi}=135\mev,\,M_{K^{0}}=495\mev,\,M_{\eta}=548\mev,\, M_{\eta^{\prime}}=958\mev$. 

From solving the mass gap equations of Eq.~(\ref{eq:Effmass}) for (2+1)-flavor NJL models, the dynamical quark masses as functions of the quark chemical potential can be obtained, as shown in Fig.~\ref{fig:mass_rho}. When $\mu_f^*<M_f$, the dynamical quark masses retain the vacuum value, where quarks strongly interact and the system corresponds to the chirally broken phase. As the chemical potential increases, the dynamical quark masses decrease while $\mu_f^*>M_f$. Meanwhile, the quark number densities become nonzero as shown in the \black{right} panel of Fig.~\ref{fig:mass_rho}. With increasing $\alpha$, the vector interactions contributed by the Fierz transformed channels in Eq.~(\ref{eqFierzscalar}) make the dynamical mass decrease slowly and the EOS stiffer than the original NJL model EOS. This will impact the interplay between the deconfinement and chiral phase transitions, which will be discussed in detail in Section~\ref{Sec: Results}.

\subsection{QCD vacuum pressure as free parameter and the determination of chiral phase transition}  \label{Subsec: QCD vacuum pressure}
At finite chemical potential and zero temperature, the pressure for quark matter can be strictly proved with the functional path integrals of QCD~\cite{2008PhRvD..78e4001Z,2008IJMPA..23.3591Z}:
\begin{equation}
P(\mu; M)=P(\mu=0; M)+\int_{0}^{\mu} d \mu^{\prime} \rho\left(\mu^{\prime}\right) , \label{eq: njLPressure}
\end{equation}
in which $M$ is a solution of the gap equation shown before.
The energy density and pressure of the system have the thermodynamic relation of
\begin{equation}
\varepsilon
=-P+\sum_{i=u, d, s, e} \mu_{i} \rho_{i}\left(\mu_{i}\right)\ , \label{eq:NJL eos}
\end{equation}
with the $\beta$-stable and charge neutrality conditions.
Due to the non-perturbative difficulty of calculating the vacuum pressure $P(\mu=0; M)$ at $\mu=0$ in Eq.~(\ref{eq: njLPressure}) from the first-principles QCD, currently, one has to make use of various effective QCD models. In NJL-type models, people usually choose the trivial vacuum as $P(\mu=0; m)$, and evaluate the vacuum pressure difference between the trivial vacuum $P(\mu=0; m)$ and the spontaneous symmetry breaking non-trivial vacuum $P(\mu=0;M)$ to determine the vacuum pressure, as extensively discussed in Ref.~\cite{2022PhRvD.105l3004Y}.
Nevertheless, this procedure to determine the vacuum pressure is unsatisfactory, due to the lack of confinement at vanishing density.
Therefore, in this study, we take $P(\mu=0; M)$ as a phenomenological free parameter corresponding to $-B$ (vacuum bag constant)~\cite{2020PhRvD.101f3023L,2022PhRvD.105l3004Y,2024ApJ...966....3Y,2012ApJ...759...57L}, which preserves the confinement of quarks and give the model high flexibility to construct the hybrid EOS.

Following the viewpoint of non-vanishing vacuum pressure, Ref.~\cite{2008PhRvD..77f3004P} fix a bag constant for the deconfinement to occur along with the chiral symmetry restoration, leading to a significant change in the EOS with respect to the conventional procedure. With increasing attention to the connection between the chiral and deconfinement phase transitions, it has been suggested that the deconfinement and chiral phase transitions split at the critical endpoint. Consequently, a confined but chiral symmetric phase, called the quarkyonic phase~\cite{2007NuPhA.796...83M,2011PhLB..696...58H,2011PhRvD..84c4028S}, may exist in the high baryon density region, based on arguments valid in the large $N_c$ limit. In the present study, we aim to explore the possibility of having chiral restoration and deconfinement occurring at different densities, instead of modeling the quarkyonic matter.

In the following, assuming there is a sharp interface (Maxwell construction) between the bulk hadronic phase and the quark phase due to the disfavored mixed phase by the surface tension and electrostatic energy costs~\cite{2001PhRvD..64g4017A}, the deconfinement phase transition occurs at a certain baryon chemical potential at which the pressures of these phases are equal, which we define this point as $\mu_{\mathrm{de}}$ for convenience. The NJL model can reproduce the fundamental aspect of spontaneous chiral symmetry breaking in QCD \black{at the quark level}, in which the chiral condensate \black{serves as the good} order parameter for chiral phase transition, \black{as described by Landau’s phase transition theory. This helps} us to facilitate a clearer understanding of chiral phase transition. 
Utilizing the chiral susceptibility, \black{which measures how the chiral condensate (the order parameter) responds to changes in the bare quark mass, defined as bellows:}
\begin{equation}
\chi=\frac{\partial \sigma}{\partial m}\ , \label{eq:chiralsus}
\end{equation}
we can obtain the exact position defined as $\mu_{\chi}$ at which the chiral condensate exhibits the maximum change for several NJL model parameter sets. 
We note that in the PDM model, on the other hand, chiral symmetry restoration occurs through a distinct physical mechanism involving hadronic degrees of freedom. The $\sigma$ condensate contributes to chiral-variant components of nuclear mass in Eq.~(\ref{Eq: PDM mass}), as the chiral condensate $\sigma$ decreases with increasing density, the masses of the parity partners become degenerate. This degeneracy signals the restoration of chiral symmetry at the hadronic level.
In this study, we focus on the interplay between chiral symmetry restoration and the deconfinement transition. Since chiral symmetry breaking is an intrinsic property of QCD at the quark level, it is natural to adopt a model that directly describes quark dynamics. In this regard, the NJL model provides a well-defined order parameter, enabling us to precisely identify the chiral restoration point $\mu_{\chi}$, which is essential for a systematic investigation of the split $\Delta = \mu_{\text{de}} - \mu_{\chi}$. Moreover, the vacuum pressure parameter $B$ in the NJL model allows us to systematically vary the deconfinement transition point while keeping the chiral dynamics fixed, thereby facilitating an efficient exploration of the relevant parameter space.

The chiral susceptibility of the system is shown in Fig.~\ref{fig:sus}. The finite and smooth peak of the chiral susceptibility manifests that chiral phase transition is a crossover. The threshold of the quark chemical potential which corresponds to the maximum $\mu_{\chi}$ is pushed to higher values due to the strong vector interactions at large $\alpha$, which is indicated with black stars at $\mu_q=301.0 \mev$, $\mu_q= 307.6 \mev$, $\mu_q=385.0\mev$, and $\mu_q=318.7 \mev$ for $\alpha=0.00$, $\alpha=0.60$, $\alpha=0.80$, and $\alpha=0.95$, respectively. A quite larger $\alpha$ contributes to stronger vector interactions and thus influences the chiral susceptibility significantly. 

\section{Hybrid star structure and stable parameter space with quark core} \label{Sec: Results}
In this section, we have solved the Tolman-Oppenheimer-Volkoff (TOV) equations for spherically symmetric and static stars in order to investigate the influence of the exchange channels weighting by parameter $\alpha$, the pivotal role of vacuum pressure $B$ and the important chiral invariant mass $m_0$ within PDM on the competition between the confinement phase transition and chiral phase transition, as well as on the maximum mass of hybrid stars. For the neutron star crust, we use the BPS EOS~\cite{1971ApJ...170..299B}, where the energy density of the region ranges from $1.0317 \times 10^4 \rm g/cm^3$ to $4.3 \times 10^{11} \rm g/cm^3$. The outer core is calculated within the PDM, and the possible quark core is described by the NJL model.

\subsection{Interplay between the deconfinement and chiral phase transitions}\label{Sec: Hybrid star structure}

\begin{figure}
\centering
\includegraphics[width=0.49\textwidth]{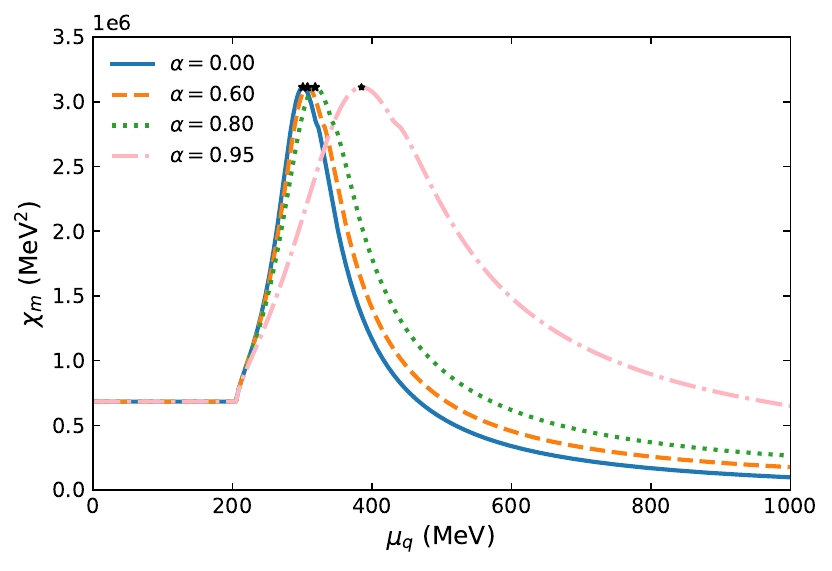}
\caption{Chiral susceptibility as a function of quark chemical potential at zero temperature for $\alpha=0.00$, $\alpha=0.60$, $\alpha=0.80$ and $\alpha=0.95$, highlighting the chiral phase transition is a crossover. The maximum change of the chiral condensates within NJL models is indicated with black stars at $\mu_q=301.0 \mev$, $\mu_q= 307.6 \mev$, $\mu_q=385.0\mev$, and $\mu_q=318.7\mev$ respectively. }\label{fig:sus}
\end{figure}
\begin{figure}[H]
\centering
\includegraphics[width=0.49\textwidth]{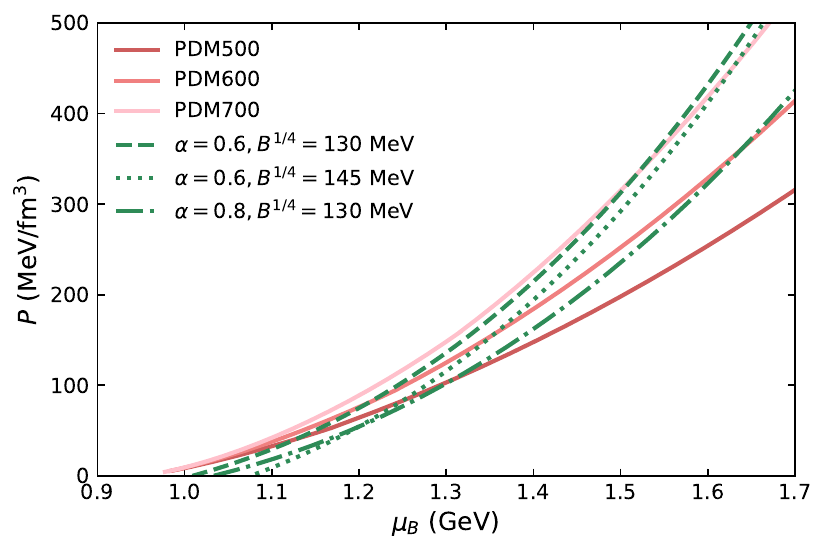}
\caption{Pressure as a function of baryon chemical potential for hadronic matter and strange quark matter. The green curves correspond to the results from the present modified NJL model, while red curves present PDM model calculations for hadronic matter with chiral invariant mass for $m_0=500\mev$, $m_0=600\mev$, and $m_0=700\mev$, respectively.}\label{fig:Pmu}
\end{figure}

We illustrate the pressure as functions of baryon chemical potential for different phases in Fig.~\ref{fig:Pmu}. In this figure, $\mathrm{PDM500}$ indicates the PDM with chiral invariant mass for $500\mev$. It can be seen that a smaller invariant mass $m_0$ in the PDM leads to high fermi energy at the same pressure, which contributes to a stiffer hadronic matter EOS. As already known from our previous works~\cite{2015PhRvC..92b5201M,2017PhRvC..95e9903M,2021PhRvC.103d5205M,2021PhRvC.104f5201M,2022PhRvC.106f5205G,2023Symm...15..745M,2024PhRvC.109f5807G}, large $m_0$ cannot produce a stiff enough pure hadronic matter EOS to support the current observations of massive compact stars. It has been established since that the first-order phase transition leads to a softening of the EOS and lower maximum masses of the sequences of hybrid stars compared to their purely hadronic counterparts. In this study, we focus on the PDM with $m_0=500\mev$ and $m_0=600\mev$ for our present study. Since the impact on the chiral phase transition is inconspicuous for the small values of $\alpha$ as shown in Fig.~\ref{fig:sus},
in the modified NJL models, we select two typical cases of $\alpha$: $\alpha=0.6$ and $\alpha=0.8$, for discussion. With the same vacuum pressure $B^{1/4}=130\mev$, the strong repulsive interactions in the exchange channels for large $\alpha$ can modify the deconfinement phase transition density $\mu_{\rm de}$ to the high value. Because a stiffer quark matter EOS leads to a higher $\mu_{\rm de}$ (and thus a higher baryon chemical potential), requiring a higher pressure to overcome the confining effect of the hadronic matter, which can be seen clearly from a comparison of $\mu_{\rm de}$. This resembles an increase in vacuum bag constant $B$, which consequently leads to a very different hybrid EOS.

\begin{figure}[H]
\centering
\includegraphics[width=0.49\textwidth]{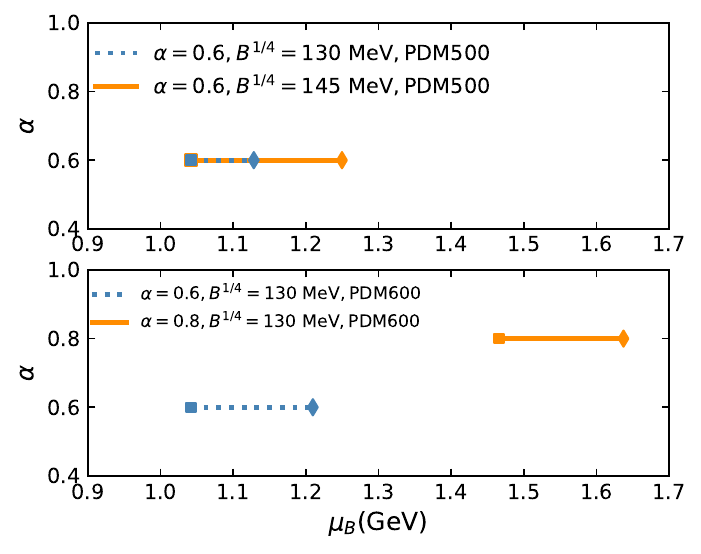}
\caption{The position of the chiral phase transition and the deconfinement phase transition, indicated with colored squares and stars respectively, as functions of baryon chemical potential for hybrid EOS adopting $\alpha=0.0$, $B^{1/4}=150 \mev$; $\alpha=0.6$, $B^{1/4}=145\mev$, and $\alpha=0.8$ $B^{1/4}=150 \mev$ within NJL model for quark phase and $m_0=500\mev$ or $m_0=600\mev$ within PDM for hadronic phase.}\label{fig:ciralanddecon}
\end{figure}

To illustrate the interplay between deconfinement and chiral symmetry restoration, we present the corresponding baryon chemical potential $\mu_B$ at which the phase transition happens with several typical parameter sets as shown in Fig.~\ref{fig:ciralanddecon}. The maximum change of chiral condensate occurring at $\mu_{\chi}$, obtained from NJL model, is indicated by squares, while the deconfinement occurs at the chemical potential $\mu_{\rm de}$ represented by rhombuses. We define the regime between $\mu_{\chi}$ and $\mu_{\mathrm{de}}$ as $\Delta=\mu_{\mathrm{de}} - \mu_{\chi}$ for convenience. Keeping the other parameters unchanged, a stiffer hadronic matter EOS with $m_0=500\mev$ narrows the chiral symmetry restored but confined phase, and makes the $\mu_{\rm de}$ to low values of the baryon chemical potential. This results in the advanced appearance of the strange quark matter and, subsequently a softer hybrid matter EOS. \black{Large vacuum pressure $B$ promotes the $\mu_{\mathrm{de}}$ to a high baryon chemical potential, and then enlarges the $\Delta$'s region with a typical hadronic matter EOS. In the case with fixed $\alpha$ (e.g., $\alpha=0.6$) for $\rm PDM500$, we obtain $\Delta=0.086\gev$ and $\Delta=0.208\gev$ for $B^{1/4}=130\mev$ and $B^{1/4}=145\mev$, respectively, as shown in the upper panel of Fig.~\ref{fig:ciralanddecon}. The effect of the bag constant on $\Delta$ is straightforward. Since vacuum pressure $B$ in the NJL model only affects the position of $\mu_{\rm de}$, increasing $B$ pushes $\mu_{\rm de}$ to a high value.} 

\begin{figure}[H]
\centering
\includegraphics[width=0.49\textwidth]{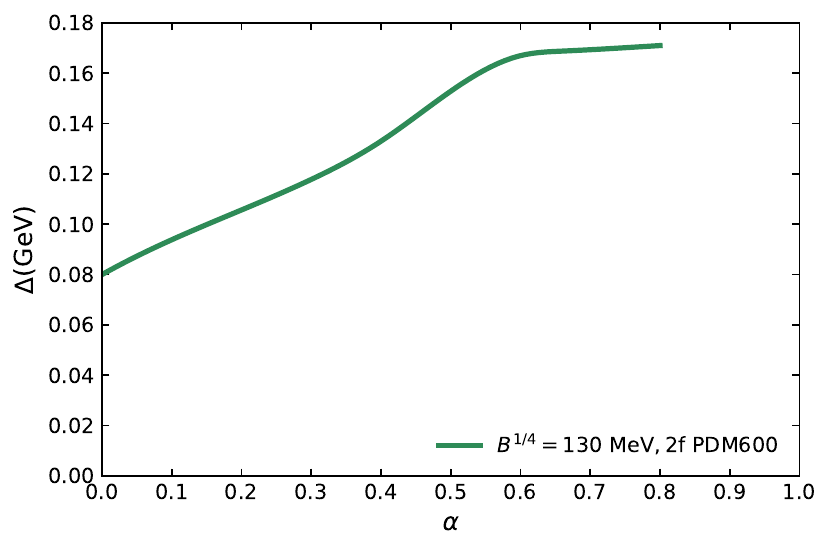}
\caption{The split between the chiral phase transition and the deconfinement phase transition($\Delta=\mu_{\rm de}- \mu_{\chi}$) as a function of $\alpha$ for hadronic matter with $\rm PDM600$ and quark matter EOS at $B^{1/4}=130\mev$.}\label{fig:Delta}
\end{figure}

\black{The crucial importance of the weighting factor $\alpha$, which manifests the contribution of the exchange channel, can be found from the comparison of the following two cases: When keeping other parameters unchanged, both $\mu_{\chi}$ and $\mu_{\mathrm{de}}$ are pushed to higher chemical potentials obsviously, while enlarging the regime between $\mu_{\chi}$ and $\mu_{\mathrm{de}}$, as clearly shown in the lower panel of Fig.~\ref{fig:ciralanddecon}. For $\alpha=0.6$, the chiral symmetric but confined regime is $\Delta=0.168\gev$, while for $\alpha=0.8$, this region is $\Delta=0.171\gev$. The effects relating to $\alpha$ can be understood from its repulsive nature at finite chemical potential. 
In Fig.~\ref{fig:Delta} the split between the chiral phase transition and the deconfinement phase transition $\Delta$ shows a nonmonotonic increasing behavior as a function of $\alpha$. This trend arises from the growing contribution of strong vector interactions with increasing $\alpha$. The vacuum bag constant $B$ competes with $\alpha$ resulting in the nearly horizontal behaviors, which in turn affects the hybrid star configuration. A discussion of this effect will be provided later.}


\begin{figure*}[htbp]\centering
\includegraphics[width=0.9\hsize]{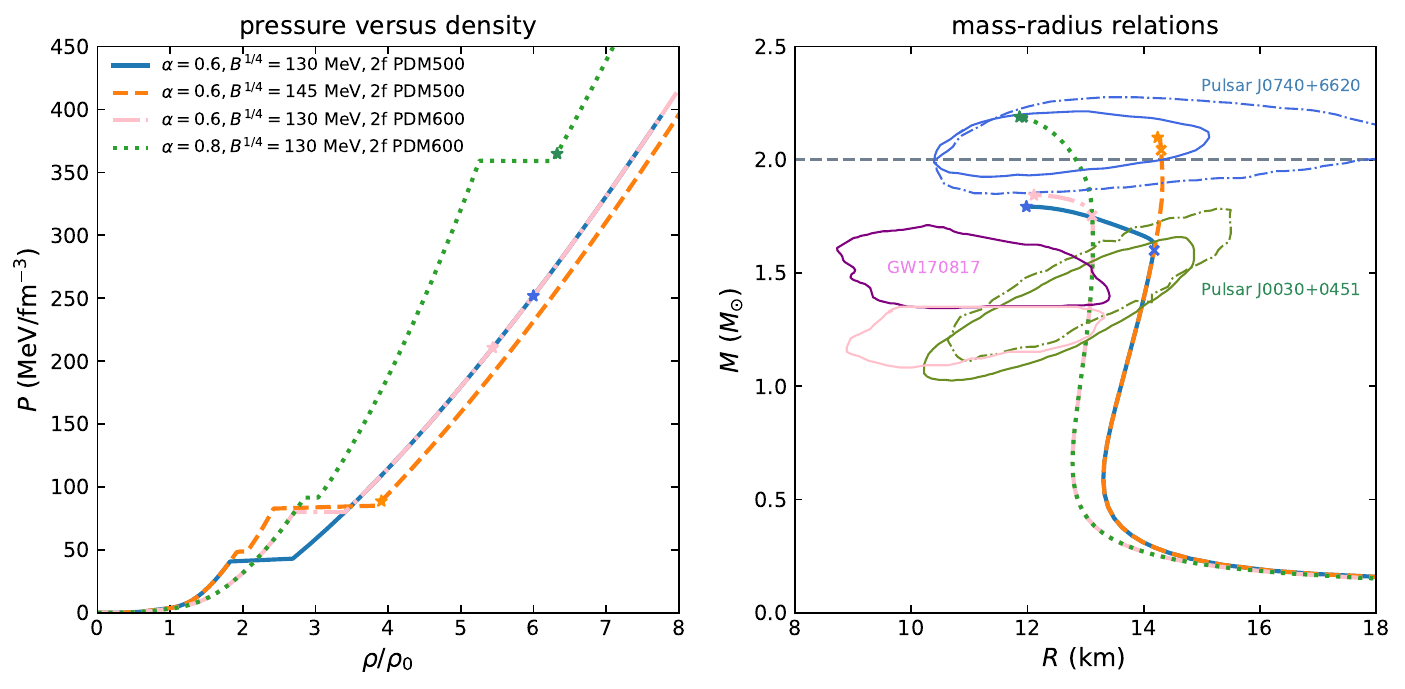}
\includegraphics[width=0.47\hsize]{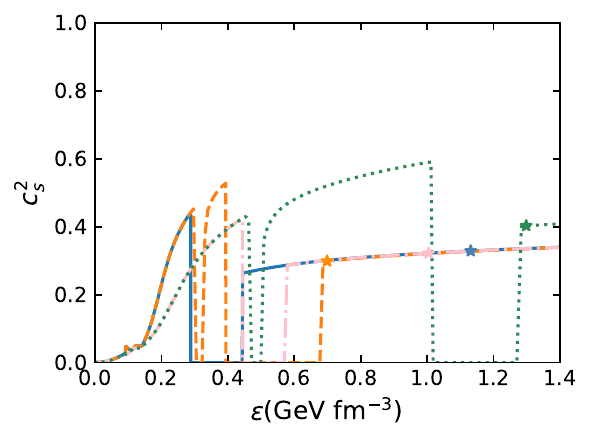}
\includegraphics[width=0.47\hsize]{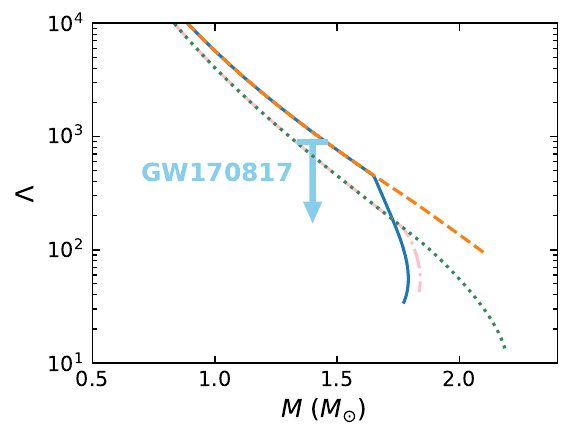}
\caption{\black{Upper left panel}: Pressure of hybrid star matter versus density (in units of the nuclear saturation density $\rho_0$) for various choices of parameter sets. The central densities $\rho_c$ of the corresponding maximum-mass hybrid stars are indicated with colored stars, respectively.
\black{Upper right panel}: The mass-radius relations for the corresponding parameter sets are shown. The maximum masses are marked with colored stars, while the crosses represent the beginning of the contribution of quark matter to the masses. The obtained quark cores are $0.24\msun$($\alpha=0.6, B^{1/4}=130\mev, \rm PDM500$), $0.06\msun$ ($\alpha=0.6, B^{1/4}=145\mev, \rm PDM500$), $0.09\msun$ ($\alpha=0.6, B^{1/4}=130\mev, \rm PDM600$), and $0.01\msun$ ($\alpha=0.8, B^{1/4}=130\mev, \rm PDM500$), respectively. The available mass-radius constraints from the NICER mission (PSR J0030 + 0451 ~\cite{2019ApJ...887L..24M,2019ApJ...887L..21R}  and PSR J0740 + 6620 ~\cite{2021ApJ...918L..28M,2021ApJ...918L..27R}) and the binary tidal deformability constraint from LIGO/Virgo (GW170817~\cite{2017PhRvL.119p1101A,2018PhRvL.121p1101A}), at the 90$\%$ confidence level, are also shown together. \black{Lower left panel: Sound speed squared $c_s^2$ (in units of the speed of light squared $c^2$ versus energy density for the corresponding parameter sets.  Lower right panel: Dimensionless tidal deformability $\Lambda$ as functions of stellar mass $M(M_\odot)$ are shown, together with the constraint from the GW170817 event, which requires $\Lambda<800$~\cite{2017PhRvL.119p1101A,2018PhRvL.121p1101A}.}}
\label{fig:eosMR}
\end{figure*}

\black{As discussed before}, large vacuum pressure $B$ promotes the $\mu_{\mathrm{de}}$ to a high baryon chemical potential. This property can also be found in the pressure-versus-density relations shown in the \black{upper} left panel of Fig.~\ref{fig:eosMR}. In this figure, the colored stars correspond to the central density of the maximum mass. It can be seen that increasing the vacuum pressure $B$ leads to a large density jump between these two phases, creating a massive hybrid star, while reducing the size of the quark core. The effect of $B$ is straightforward since, by definition, it is the energy excess between the perturbative and the nonperturbative vacuum, and there is a large contribution to the hybrid EOS of the chirally symmetric but confined regime for larger $B$, subsequently allowing the compact star to accumulate more mass before converting to the relatively softer quark matter phase, ultimately leading to a larger maximum mass. 
\black{The strong vector interactions in the exchange channels with large $\alpha$ in the modified NJL model produce stiff enough hybrid EOS to support massive compact star observations.} Nevertheless, as the onset of the deconfinement phase transition is delayed, the possibility that quark cores can exist in stellar objects decreases. Meanwhile, the lower onsets of the $\mu_{\mathrm{de}}$ allow the pure strange quark matter core, but soften the hybrid matter EOS. Details on the macroscopic properties of hybrid stars are given in Fig.~\ref{fig:eosMR}.

\black{We additionally calculate another important quantity of stellar matter: the sound speed, defined as $c_s = \sqrt{{\rm d}P/{\rm d}\epsilon}$, as shown in the lower left panel of Fig.~\ref{fig:eosMR}. The sound velocity exhibits a nearly monotonically increasing behavior for nuclear matter at lower energy densities. Before $\mu_{\rm de}$, the chiral restoration in the PDM leads to a small discontinuity in the behavior of $c_s^2$. During the first-order deconfinement phase transition, the sharp jump in energy density at constant pressure results in $c_s^2 = 0$. For quark matter, variations in the vacuum pressure do not alter the values of the sound speed but significantly shift the position of $\mu_{\rm de}$. As shown in Fig.~\ref{fig:eosMR}, increasing the parameter $\alpha$ makes the quark matter equation of state stiffer, corresponding to a larger $c_s^2$.}

\subsection{Mass-radius relations}
The features discussed above are elucidated further by examining the gravitational mass-radius relations under varying vacuum bag constants $B$ and Fierz transformed vector interactions characterized by 
$\alpha$. Fig.~\ref{fig:eosMR} presents the mass $M(\msun)$ of hybrid stars as a function of radius $R$ and the corresponding center densities $\rho_c$ for several selected cases. In this analysis, we primarily consider the parameters influencing the stability of hybrid stars with quark cores. Constraints from astronomical observations will be addressed in subsequent discussions.

By comparing the cases of $\rm PDM 500$ and $\rm PDM 600$ with quark matter EOS for $\alpha=0.6, B^{1/4}=130\mev$, it is evident that the hadronic matter EOS primarily determines the radius of the hybrid star. Increasing the invariant mass $m_0$ softens the hadronic matter EOS, resulting in a relatively stiffer hybrid star EOS that can support more massive compact stars, due to the delayed deconfinement phase transition expands the region of quarkyonic matter. The softening of the hadronic EOS significantly reduces the radius. As shown in the right panel of Fig.~\ref{fig:eosMR}, the radius of a $1.4 M_{\odot}$ hybrid star shrinks by about $1 \rm km$ when $m_0$ is increased from $500 \mev$ to $600\mev$. This adjustment brings the mass-radius relation into better agreement with both the LIGO/Virgo and NICER constraints, as well as the constraint from the PSR J0437-4715~\cite{2024ApJ...971L..20C,2024ApJ...971L..19R}, which has a $1.4 \ M_{\odot}$ and a relatively small radius compared to $\rm PSR J0030+0451$. A soft hadronic matter EOS also influences the central density of hybrid stars, typically suggesting a lower central density. This is also because $\mu_{\rm de}$ is pushed to higher densities, leading to a stiffer hybrid star EOS.

A larger value of $\alpha$, signifying stronger vector interactions, results in a stiffer quark matter EOS and higher deconfinement density, allowing the neutron star to sustain a more massive hadronic matter shell before transitioning to entering the quark matter phase. 
Additionally, the stiffer quark matter EOS provides greater pressure support against gravitational collapse, enabling the hybrid star to support a higher maximum mass. This effect is clearly demonstrated in the right panel of Fig.~\ref{fig:eosMR} by comparing the green curve with the pink one, which corresponds to $\alpha = 0.8$ and $\alpha = 0.6$, respectively. The hybrid star with $\alpha = 0.8$ achieves a higher maximum mass around $2.2\msun$ compared to the one with $\alpha = 0.6$ $(\sim 1.8\msun)$. 

\begin{figure*}
\centering
{\includegraphics[width=0.49\textwidth]{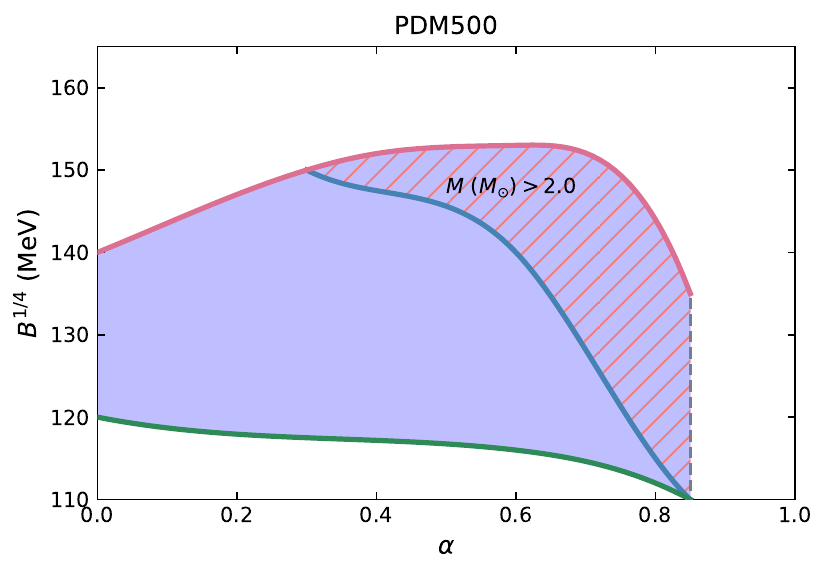}}
{\includegraphics[width=0.49\textwidth]{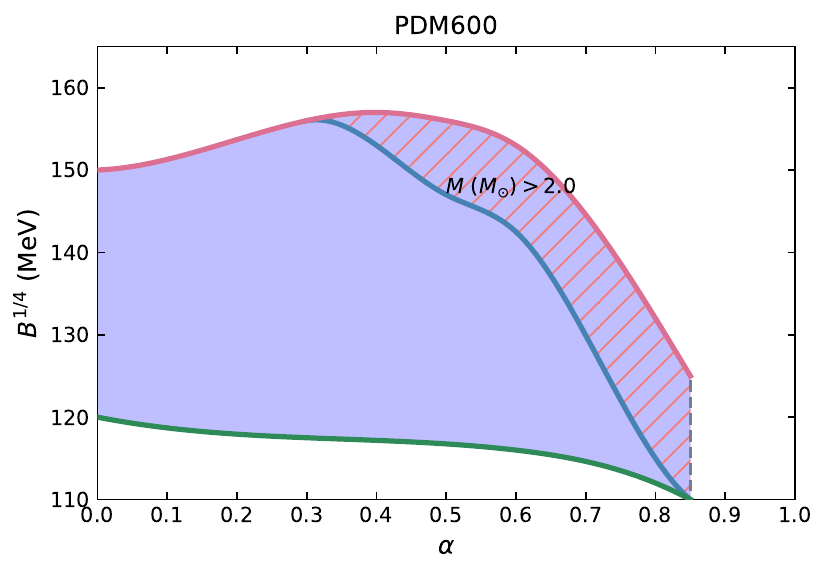}} 
\caption{Parameter spaces for stable hybrid star configurations are shown with blue shaded regions. The green solid lines correspond to the minimum vacuum pressure in the NJL model needed to be able to construct a stable hybrid star for each $\alpha$. The purple solid lines indicate the maximum vacuum pressure as a function of $\alpha$, below which the stable hybrid star has a quark core. The blue solid lines show the parameter sets that lead to $2\msun$ hybrid stars. The skewed shaded region represents the parameter spaces that satisfy current observations of the maximum mass.}
\label{fig:parameterSpace}
\end{figure*}
%
It is important to note that if the value of $B$ is too large, \black{the configuration of a hybrid star with a quark core becomes unstable due to the imbalance of pressure and gravity}, imposing constraints on the acceptable range of $B$ values. The presence of a quark matter core is crucial for the existence of a hybrid star, as it provides the necessary pressure support to sustain the star against gravitational collapse. If the bag constant is too large, the quark matter EOS becomes too soft, and the pressure support from the quark matter core becomes insufficient to counteract the gravitational force, resulting in an unstable hybrid star configuration. 

\black{The binary neutron star merger event GW170817 provides a constraint on dimensionless tidal deformability $\Lambda<800$ for a \(1.4\,M_{\odot}\) compact star~\cite{2017PhRvL.119p1101A,2018PhRvL.121p1101A}. In Fig.~\ref{fig:eosMR}, we also present direct calculations of \(\Lambda\) for hybrid stars containing quark matter cores, corresponding to the selected parameter sets. To linear order, the tidal deformability $\lambda$, which characterizes the response of a neutron star to an external tidal field, is related to the $l=2$ dimensionless tidal Love number $k_2$ as
\begin{equation}
\begin{aligned}
\lambda &=\frac{2}{3} k_2 R^5\ , \\
\Lambda &= \lambda/M^5\ ,
\end{aligned}
\end{equation}
using geometric units with $G =c= 1$. Here, $\Lambda$ is the dimensionless tidal deformability, with $M$ and $R$ representing the mass and radius of the star, respectively. Following the method developed in Refs.~\cite{2008ApJ...677.1216H,2009PhRvD..80h4035D,2010PhRvD..82b4016P,2020PhRvD.102b8501T}, we solve the TOV equations simultaneously with the perturbation equations to calculate the tidal deformability for a given EOS.  Considering the finite energy density discontinuity in the EOS, there exists a jump in the energy density at a constant pressure $P_{\rm tr}$. Therefore, $\mathrm{d}\epsilon/{\mathrm{d} P}$ exhibits a delta-function-like behavior across the discontinuity, and we can introduce an additional term to properly handle the jump, as discussed in detail in Refs.~\cite{2009PhRvD..80h4035D,2010PhRvD..82b4016P,2019PhRvD..99h3014H,2020PhRvD.102b8501T,2021PhRvD.104d3002L,2025PhRvD.112b3019Y}. The resulting behavior of the dimensionless tidal deformability $\Lambda$ as a function of the stellar mass is shown in Fig.~\ref{fig:eosMR}. The results indicate that the mass–radius relations derived from models consistent with the LIGO/Virgo constraint at the $90\%$ confidence level also yield tidal deformabilities within the allowed range. The behavior of $\Lambda$ exhibits a distinct inflection point when this characteristic also appears in the corresponding mass–radius curve. The tidal deformability constraint from GW170817 favors larger values of $m_0=600\mev$, which correspond to a slightly softer hadronic EOS.}

From the above discussions, we conclude that a softer hadronic matter EOS at intermediate densities, combined with a sufficiently stiff quark matter EOS at high densities, enables the hybrid star with quark core to reconcile well with the radius constraints for $1.4\msun$ stars and the high mass constraints from $\rm PSR~J0740+6620$. This configuration supports a hybrid star model that meets both the observed radii of typical neutron stars and the observed masses of the most massive known neutron stars. 

\subsection{Stable parameter space with quark core}
In this part, we show the obtained parameter space for stable hybrid stars in the $B^{1/4}-\alpha$ plane, as shown in Fig.~\ref{fig:parameterSpace}, for the two case of PDM with different chiral invariant masses: $m_0=500 \mev$ and $m_0=600\mev$.
In each case, the purple line represents the upper limit of the value of $B^{1/4}$ that allows \black{the stable hybrid star configuration with a quark core at its center} for a given value of $\alpha$. The minimum value of the vacuum pressure $B^{1/4}$, indicated with the green solid lines, should satisfy that the deconfinement phase transition cannot be shifted into a density regime where the chiral symmetry restoration does not occur. This minimum $B^{1/4}$ should also ensure that the quark matter EOS can construct a hybrid EOS with the PDM-calculated hadronic matter EOS. It is evident that the blue shaded regions correspond to the defined parameter spaces for stable hybrid star configurations with quark cores. Recent astrophysical observations of massive neutron stars impose tight constraints on the allowed parameter space. The skewed shaded regions bounded by the purple and blue lines in Fig.~\ref{fig:parameterSpace} denote the parameter spaces that satisfy the current observational constraints on the maximum mass of $2 M_{\odot}$. When $\alpha>0.85$, the pressure of the quark matter system will be higher than hadronic matter, which supports the possible existence of an inverted hybrid star~\cite{2023PhRvD.108f3012Z,2024PhRvD.109f3020Z,2024arXiv240706410N}.

Focusing on the softer hadronic EOS with $\rm PDM600$, in this case the slight decrease of the minimum vacuum pressure $B$ results from the stiffer quark matter EOS for larger $\alpha$. This can be understood from the quark number density relation in Fig.~\ref{fig:mass_rho}. Because the threshold where the $\rho_s$ appears is pushed to a higher chemical potential by the strong vector repulsion at larger $\alpha$. Then the $\mu_B$ at $P=0$ is pushed to a high value (see Fig.~\ref{fig:Pmu}), thus the quark matter EOS with large $\alpha$ does not need large $B$ to ensure that the hadronic and quark EOS can construct a hybrid star configuration. The relationship between the minimum $B$ and $\alpha$ does not change when the stiffness of the hadronic matter EOS is varied, because the property of the PDM is the same at lower chemical potentials as shown in Fig.~\ref{fig:Pmu}.
The maximum vacuum pressures $B$ in the NJL model first increase slightly, then decrease with a stiff quark matter EOS at large $\alpha$. From the discussion above in Fig.~\ref{fig:eosMR}, the results tell us that the quark matter EOS should be stiff enough to allow the neutron star to have a quark core. Both increasing $\alpha$ and $B$ can lead to a larger maximum mass of the hybrid star, but a smaller quark core. As $\alpha$ becomes larger and larger, giving the system strong vector interactions, the maximum $B$ should be reduced to ensure a stable hybrid star with $M > 2 \msun$.

\section{Conclusions and summary}\label{sec:summary}
In this work, we perform a systematic study of hybrid star configurations by employing a PDM-derived hadronic EOS and a modified NJL Model for three-flavor quark matter. The NJL model Lagrangian incorporates scalar interactions, ’t Hooft interactions, and Fierz-transformed interacting channels. The parameter $\alpha$ is treated as a free parameter from $0$ to $1$. We account for the possibility of a distinct separation between the deconfinement and the chiral phase transitions and explore their interplay on the stability of hybrid stars, assuming a first-order deconfinement phase transition. This separation can be adjusted by defining the vacuum pressure as a free parameter $B$ (a bag constant) in the NJL model. Our analysis has shown that the interplay between the non-perturbative interactions, as well as the magnitude of the split between the chiral and deconfinement phase transitions, have significant implications for the structure and maximum mass of hybrid stars. The chiral invariant mass $m_0$ in the PDM plays a crucial role in determining the stiffness of the hadronic EOS, while the parameters $\alpha$ and $B$ in the NJL model govern the stiffness of the quark matter EOS. The interplay between the non-perturbative interactions within PDM and NJL model controls the location of the deconfinement phase transition.

We find that a relatively large invariant mass $m_0$ around 600 MeV is necessary to ensure that the obtained radius can satisfy the constraints on the tidal deformability of a star $\sim 1.4\msun$ in the GW170817 event by the LIGO-Virgo Collaboration. That is, the intermediate density EOS should be moderately soft. And the extended chiral symmetry restored but confined phase leads to a stiff hybrid star EOS, but reduced quark matter core. Furthermore, to be consistent with the maximum mass constraints, the quark matter EOS should be stiff enough. Thus, the parameter $\alpha$ in the modified NJL model with strong vector interactions in the quark matter system should be large enough. The vacuum bag constant $B$ plays a pivotal role in changing the position of $\mu_{\rm de}$. As we increase the value of $B$, the hybrid stars have a larger maximum mass, but a smaller quark core with less stability. By comparing our theoretical results with observational constraints from gravitational wave measurements (LIGO/Virgo) and pulsar observations (NICER), we have identified the parameter spaces that lead to stable hybrid star configurations consistent with current data. Our analysis suggests that a larger split between the chiral and deconfinement phase transitions is favored by the observations, while resulting in a smaller quark core. The maximum mass with a quark core we obtained is $\sim 2.2\msun$ for $B^{1/4}=125\mev$, $\alpha=0.85$ at $\rm PMD 600$.

Our study highlights the importance of considering the interplay between the hadronic and quark matter EOSs, as well as the magnitude of the split between the chiral and deconfinement phase transitions, in the modeling hybrid stars. By incorporating both chiral symmetry breaking and deconfinement effects, and systematically exploring the parameter space, we develop a more comprehensive framework for understanding the properties of dense matter under extreme conditions.

\medskip
\acknowledgments
We thank Chengjun Xia, Yonghui Xia, Bing-Jun Zuo, Cheng-Ming Li, and Zhi-Qiang Miao for the helpful discussions. This work is supported by the National SKA Program of China (2020SKA0120100), the National Natural Science Foundation of China (Nos. 12003047, 12133003), and the Strategic Priority Research Program of the Chinese Academy of Sciences (No. XDB0550300). Wen-Li Yuan is supported by the Special Funds of the National Natural Science Foundation of China (Grant No.~12447171) and the China Postdoctoral Science Foundation (Grant No.~2025M773418). Bikai Gao is supported in part by JSPS KAKENHI Grant Nos.~20K03927, 23H05439, 24K07045, and JST SPRING, Grant No. JPMJSP2125.  Bikai Gao would like to take this opportunity to thank the “Interdisciplinary Frontier Next-Generation Researcher Program of the Tokai Higher Education and Research System.”

\end{document}